# Nanosize confinement induced enhancement of spontaneous polarization in a ferroelectric nanowire

M. Q. Cai[1, 2, 3], Y. Zheng[1], B. Wang[1] and G. W. Yang[1a)]

[1]*State Key Laboratory of Optoelectronic Materials and Technologies, Institute of Optoelectronic and Functional Composite Materials, Nanotechnology Research Center, School of Physics & Engineering, Zhongshan University, Guangzhou 510275, Guangdong, P. R. China*

[2]*School of Physics and Microelectronics Science and Micro–Nanotechnologies Research Centre, Hunan University, Changsha 410082, Hunan, P. R. China*

[3]*State Key Laboratory of Powder Metallurgy, Central South University, Changsha, Hunan, 410083, P. R. China*

## Abstract

We theoretically showed that the spontaneous polarization in ferroelectric (FE) nanowires (NWs) can be considerably enhanced due to the nanosize confinement by the first-principles calculations. The spontaneous polarization in a fully-relaxed $PbTiO_3$ NW with 1.8 nm diameter is 1.26 times higher than that of bulk counterpart. The tension induced by NW surface curvature counteracts the near-surface depolarizing effect and meanwhile leads to the unusual enhancement of spontaneous polarization. These findings indicated that FE NWs can be promising in the applications of nanodevices.

a) Correspondingauthor: stsygw@mail.sysu.edu.cn

Low-dimensional ferroelectrics (FEs) have attracted intense attentions due to their potential applications in leading toward miniaturized devices such as high density non-volatile ferroelectric random access memories with the large storage density [1-3]. Much experimental effort has been made recently in preparing and understanding FE nanorings [4], nanowires (NWs) [5-7], and nanoparticles [8, 9]. Theoretically, Naumov *et al.* [10-12] used a first-principles-derived effective-Hamiltonian approach to study the ferroelectricity in $BaTiO_3$ nanodots, $Pb(Zr,Ti)O_3$ nanodisks and rods, and revealed the vortex structural dipoles in these nanostructures. Geneste *et al* [13] studied the size-dependence of ferroelectric properties of $BaTiO_3$ NWs and found that the ferroelectric distortion could be recovered under appropriate tensile strain conditions by the first-principles calculations. Ferroelectricity is generally caused by atomic off-center displacements that resulting from a delicate balance between long-range (LR) Coulomb interaction and short-range (SR) covalent interaction [14]. In one-dimensional (1D) FE nanostructures, LR interaction is truncated due to the lack of periodicity and SR one is significantly modified near the surface boundary. Therefore, the depolarizing effect would come into being because of the lack of periodicity and the modified surface boundary, which causes the spontaneous polarization decreasing. [2, 10, 12]. However, Morozovska *et al.* [15] reported that the compressive stress induced by surface curvature would produce an effective tensile in the length direction of NW. In this letter, we carry out the first-principles calculations to pursue the spontaneous polarization in a fully-relaxed $PbTiO_3$ FE NWs by taking the near-surface depolarizing effect and the tension induced by surface curvature into account.

A PbTiO$_3$ NW is as shown in Figure 1. The vacuum space with about 8 nm thicknesses surrounds the nanowire, which ensures that NWs in neighboring supercells do not interact with each other and causes the surface-induced atomic relaxation and the cell-shape changes. These treatments would thus affect both local modes and local inhomogeneous strains. In order to keep be insulating and close to circular, besides PbO- or TiO$_2$- surface-terminations, the configurations of NWs are stochiometric with PbO and TiO$_2$ surface-terminations at the same time. Thus, the whole structure would be kept symmetrical corresponding to the central $z$-axis. We perform the simulations for a variety of diameters $d$ (1.4 to 4.2 nm), while the $z$-axis periodical length $h$ is chosen to be the bulk lattice constant. Other parameters used here are those of bulk PbTiO$_3$. First-principles density functional theory (DFT) with local density approximation is performed using the Vienna *ab initio* simulation package (VASP), within the projector augmented wave (PAW) pseudopotential method [16]. These calculations are performed with a 1×1×5 Monkhorst-Pack **k**-point mesh centered at Γ and a 460 eV plane-wave cutoff, both of which result in a good convergence of computed ground-state properties. The forces on atoms are calculated using the Hellmann-Feynman theorem, which are used to perform a conjugate gradient relaxation [17-19]. Structural optimizations are continued until the forces on the atoms converge to less 1 meV/Å. Moreover, the electronic polarization $P$ via $P = Z^*_{eff} U_{eff}$ of PbTiO$_3$ NWs is performed by the Berry phase method [19], where $Z^*_{eff}$ is the Born effective charge of the local mode, $U_{eff}$ is the local model of the cell to describe the ferroelectric instability.

There is a critical size of 1.4 nm in the spontaneous polarization of $PbTiO_3$ NWs as shown in Fig. 2, which further confirms the existence of the FE phase transition in NWs when their diameters are less than a definite size [10]. Importantly, a maximal value of spontaneous polarization that is about 1.26 times higher than that of bulk counterpart is obtained in the 1.8 nm $PbTiO_3$ NW. However, both the spontaneous polarizations gradually decrease with NW diameter increasing when the diameter is more or less than the critical size. Well known, the near-surface depolarizing effects associated with finite thickness of wires are strong in low-dimensional structures, which causes the considerably decrease of spontaneous polarization than that of bulk [2, 10, 12]. This conclusion clearly claims that FE nanostructures are not expected to be promising in applications of nanodevices. However, our studies show that the large enhancement of spontaneous polarization can be achieved in FE NWs.

To pursue the physical origin of the unusual enhancement of spontaneous polarization above, we carry out the Ti atom off-center displacement calculations. At the atomistic scale, the spontaneous polarization is mainly caused by the displacement of Ti atoms deviating from the centre positions (the paraelectronic phase) in the displacive FE $PbTiO_3$ [14]. In Fig. 3(a), the average displacement of $<u_{Ti}>$ is zero in the $x$ and $y$ directions, which means that the zero spontaneous polarization is in the perpendicular direction of $PbTiO_3$ NWs [10, 12]. Although there have considerable displacements for every Ti atom in the $x$ and $y$ directions, the sum whole displacive vector is zero in the parallel-to-surface direction to cause the zero polarization in Fig.

3(b). Thus, the FE ordering in 1D infinite PbTiO$_3$ NWs does not form any toroid moments [10-12], and the big off-center displacement in FE NWs should be originated from the compressive stress induced by surface curvature [15].

We propose a thermodynamic explain for our findings above. For FE NWs, the near-surface depolarizing effects expressed in terms of the extrapolation length can change the polarization near surfaces and associated properties. Meanwhile, the surface tension produces an effective radial pressure along the radial direction that would affect the Curie temperature, polarization and dielectric constant of NWs, which is similar to an effective compressive stress along the longitudinal polar axis. In general, the order parameter used to describe the transition between paraelectric and ferroelectric phases is the self polarization $\mathbf{P} = (0, 0, P)$ [20], in which $P$ is the result of the permanent atomic displacements responsible for the spontaneous polarization generated when a crystal becomes unstable. The total free energy of a FE NW as

$$F = 2\pi \int_0^h dz \int_0^R r dr \left\{ \frac{\alpha_0}{2}(T-T_{c0})P^2(r,z) - 2Q_{12}\sigma_{\text{eff}}P^2(r,z) + \frac{\beta}{4}P^4(r,z) + \frac{\gamma}{6}P^6(r,z) \right. \\ \left. + \frac{D}{2}[\nabla P(r,z)]^2 - \frac{1}{2}E_d P(r,z) - (s_{11}+s_{12})\left(\frac{\mu}{R}\right)^2 \right\} + F_s \quad (1)$$

where $\alpha_0$, $\beta$ and $\gamma$ are the expansion coefficients of the Landau free energy [21], $T_{c0}$ is the Curie-Weiss temperature of bulk material, $Q_{ij}$ is the electrostriction tensor. The effective uniform radial stress due to the surface tension $\sigma_{\text{eff}} = \mu/R$ in a NW with surface energy density $\mu$ and radius $R$ can compress NW in the transverse direction and stretch it along the polar axis $z$. In our case, the depolarization field $E_d$ along the $z$ direction is neglected because NW is infinite [10]. The last term $F_s$ accounts for the near-surface depolarizing effect contributions to the total free energy and is $F_s = D \int_{-h/2}^{h/2} \frac{2\pi R}{\lambda} P^2(r=R,z)dz$, where $\lambda$ is the extrapolation lengths on the surface of NWs.

Linear stability analysis is performed on the associated dynamic Ginsburg-Landau equation for a FE NW [22]. The transition temperature as

$$T_c = T_{c0} + \frac{4Q_{12}}{\alpha_0}\sigma_{eff} - \frac{D}{\alpha_0}k_c^2, \qquad (2)$$

where $k_c$ is the function of radius $R$ and extrapolation length $\lambda$ of NWs, which can be written as $\cot(k_c R) - k_c \lambda = 0$ [23]. $\sigma_{eff}$ is the effective stress induced by the surface tension of NWs. The approximated polarization $\bar{P}$ can be obtained by solving the following equation

$$\frac{\delta F}{\delta \bar{P}} = \alpha_0 \left[T - T_c(R,\lambda)\right]\bar{P} + \beta \bar{P}^3 + \gamma \bar{P}^5 = 0, \qquad (3)$$

$F$ is the total free energy of NWs with respect to approximated polarization. Then the approximated polarization $\bar{P}$ can be written as

$$\bar{P} = \left[\frac{-\beta \pm \sqrt{\beta^2 - 4\alpha_0 \gamma \left[T - T_c(R,\lambda)\right]}}{2\gamma}\right]^{1/2} \qquad (4)$$

According to these theories above, we attain the spontaneous polarization ($\bar{P}$) and the Curie temperature ($T_c$) of a PbTiO$_3$ NW as shown Figs. 4(a-b), respectively. Clearly, $\bar{P}$ and $T_c$ have the same trend with NWs' size changing. Importantly, there is a critical size for $\bar{P}$ and $T_c$, which indicates the existence of FE phase transition in NWs when their diameter are less than a definite size. Remarkably, with NWs' size increasing, both $\bar{P}$ and $T_c$ increase sharply to reach the maximization at about 2.0 nm, and then decrease to trend to the bulk value. Therefore, these theoretical deductions are qualitatively in agreement with our calculations. Accordingly, the surface compressive stress caused by 1D confinement produces an effective tensile in the length direction in NWs and then leads to a big off-center displacements that enhances the spontaneous polarization. Therefore, we can see that the enhanced spontaneous polarization in FE NWs is attributed to the competition between itself surface tension and near-surface depolarizing effect in Figure 4.

In summary, taking the near-surface depolarizing effect and the surface tension induced by 1D confinement into account, we have performed that there is the large enhancement of spontaneous polarization in the fully-relaxed $PbTiO_3$ NW by using first-principles calculation.

**Acknowledgments.** NSFC (50802026 and U0734004), the Ministry of Education (106126, 20070532010), and HNNSF (09jj4001) supported this work.

**Figure Captions**

Figure 1. Schematic illustrations of PbTiO$_3$ NWs viewed from (a) the [100] and (b) the [001] directions, respectively, and (c) the internal ferroelectric perovskite structure. The polarity axis is lying along the NWs length. The dark, gray, and red atoms are Pb, Ti, and O.

Figure 2. The longitudinal z-axis spontaneous polarization as a function of NW's diameter in unit of nm with the open-boundary condition. The dot line is the value of bulk polarization of PbTiO$_3$. In order to display the different FE order of the PbTiO$_3$ NW, the ratio of $P_{nanowires}/P_{bulk}$ is simultaneously presented. Note that the spontaneous polarization is different along the radial direction, and thus the calculated spontaneous polarization is an average value.

Figure 3. The average displacement (a) of Ti atoms along the $x$, $y$, and $z$ directions of PbTiO$_3$ NWs. The schematic illustration (b) of the displacements of the symmetric four Ti atoms along the $x$ and $y$ directions for the 2.6 nm PbTiO$_3$ NW ($|U_{1x}| = |U_{2x}| = |U_{2x}| = |U_{4x}|, |U_{1y}| = |U_{2y}| = |U_{3y}| = |U_{4y}|$).

Figure 4. $\bar{P}$ and $T_c$ of PbTiO$_3$ NWs as a function of radius. The black (solid) and blue (dashed) line represents the results with surface tension and without surface tension, respectively. In calculations, $\mu = 20$N/m from Ref. [24].

**Figure 1**

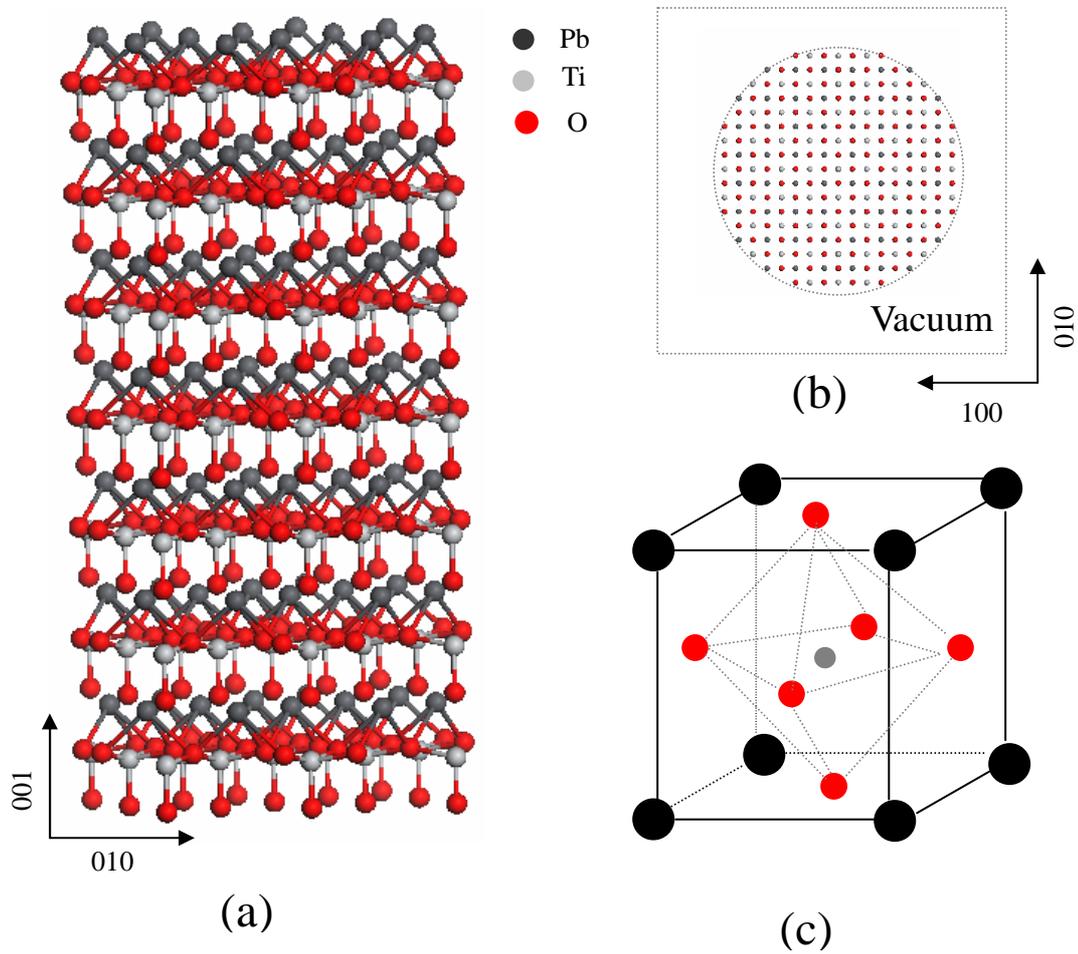

(a) (b) (c)

**Figure 2**

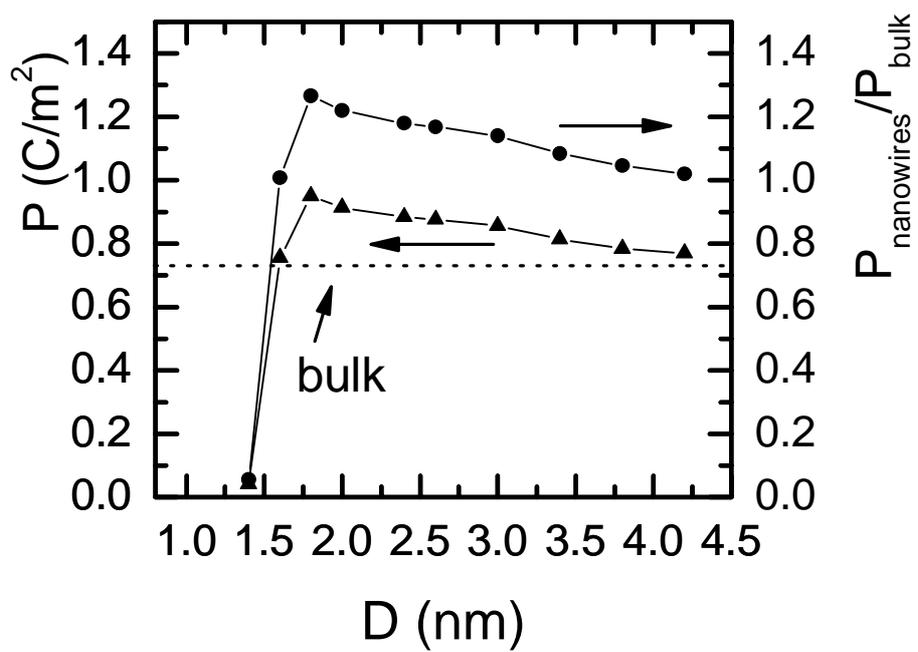

**Figure 3**

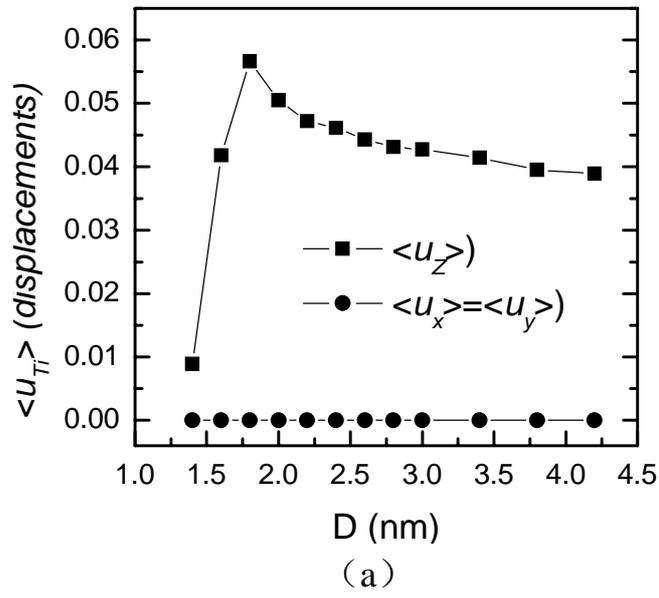

(a)

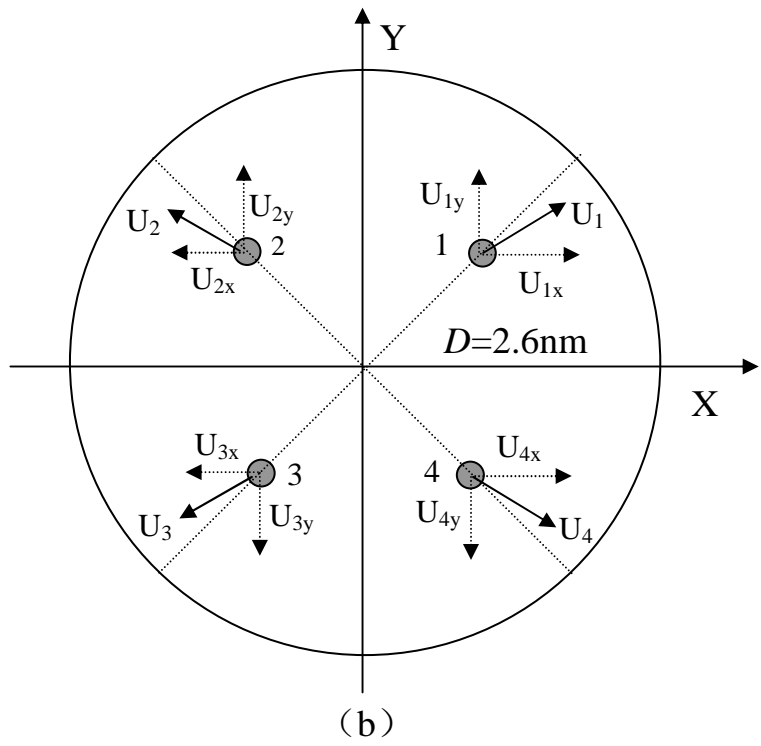

(b)

**Figure 4**

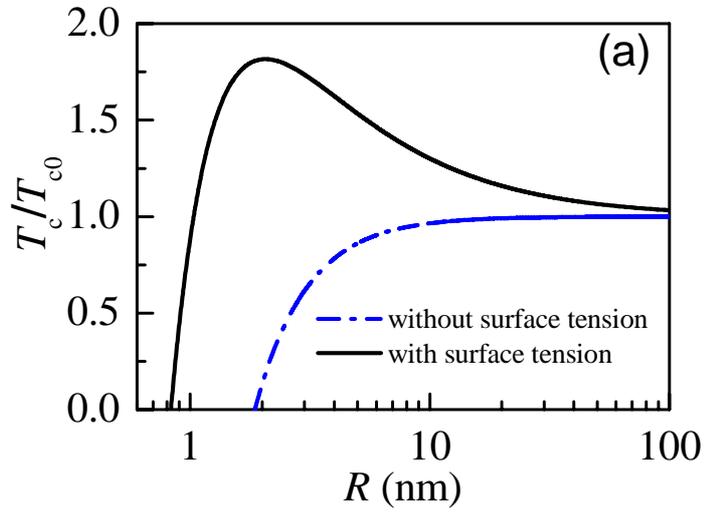

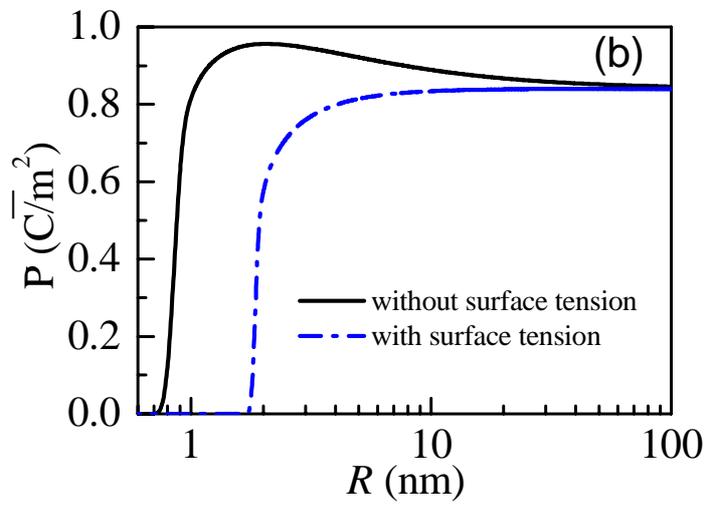